\documentclass[%
 reprint,
 amsmath,amssymb,
prc,
]{revtex4-2}

\usepackage{graphicx}
\usepackage{subcaption}
\usepackage{dcolumn}
\usepackage{bm}
\usepackage{siunitx}
\usepackage{physics}
\usepackage{wrapfig,booktabs,tabularx,tabulary}
\usepackage{threeparttable}

\newcommand{\mycomment}[1]{}

\begin{document}

\preprint{APS/123-QED}

\title{New Experimentally Observable Gamma-ray Emissions from ${}^{241}$Am Nuclear Decay}

\author{Katrina E. Koehler}
 \altaffiliation[Also at ]{Houghton University, Houghton, NY 14744}
 \email{kkoehler@lanl.gov}
\author{Michael D. Yoho}%
\altaffiliation[Currently at ]{Institute of Nuclear Sciences, Ege University, \.Izmir, T\"urkiye}
\author{Matthew H. Carpenter}
\author{Mark P. Croce}
\author{David J. Mercer}
\author{Chandler M. Smith}
\author{Aidan D. Tollefson}
\author{Duc T. Vo}
\affiliation{%
 Los Alamos National Laboratory\\
 Los Alamos, NM 87545
}%

\author{Michael A. Famiano}
\affiliation{
 Western Michigan University\\
 Kalamazoo, MI 49008
}%

\author{Caroline D. Nesaraja}
\affiliation{
Oak Ridge National Laboratory \\
Oak Ridge, TN 37830
}%

\author{Daniel T. Becker}
\author{Johnathon D. Gard}
\author{Abigail L. Wessels}
\affiliation{
University of Colorado\\
Boulder, CO 80309
}

\author{Douglas A. Bennett}
\author{J. A. B. Mates}
\author{Nathan J. Ortiz}
\author{Daniel R. Schmidt}
\author{Daniel S. Swetz}
\author{Joel N. Ullom}
\altaffiliation[Also at ]{University of Colorado, Boulder, CO 80309}
\author{Leila R. Vale}
\affiliation{
National Institute of Standards and Technology\\
Boulder, CO 80305
}


\date{\today}

\begin{abstract}
With the high resolution of microcalorimeter detectors, previously unresolvable gamma-ray lines are now clearly resolvable. A careful measurement of ${}^{241}$Am decay with a large array of gamma-ray microcalorimeters has revealed never before seen or predicted gamma lines at $207.72 \pm 0.02$~keV and $208.21 \pm 0.01$~keV. These results were made possible by new microwave-multiplexing readout to increase the array size and improved analysis algorithms to eliminate spectral artifacts. We suggest nuclear levels from which these gamma-rays might originate and calculate branching ratios for these transitions from measurements of both mixed Pu-Am standards and a pure ${}^{241}$Am source.  These results have implications for nuclear material safeguards and accounting, particularly for microcalorimeter gamma spectrometers, which are now being adopted in nuclear safeguards analytical laboratories.

\end{abstract}

\maketitle

\section{Introduction}
This paper presents two new gamma-ray emissions (spectral lines) for ${}^{241}$Am and the possible energy level candidates from which these gamma rays might originate. Plutonium and americium photon emissions around 208~keV are used for the non-destructive assay of plutonium mass by International Atomic Energy Agency (IAEA) inspectors and material control and accountability (MC\&A) technicians working in Department of Energy laboratories~\cite{PANDA}. While HPGe is unable to resolve this complex, it is very important that new branching ratios are derived for this complex of peaks to enable accurate isotopic characterizations of plutonium using microcalorimeter gamma systems.

\begin{figure}[h!]
\centering
\includegraphics[width=\columnwidth]{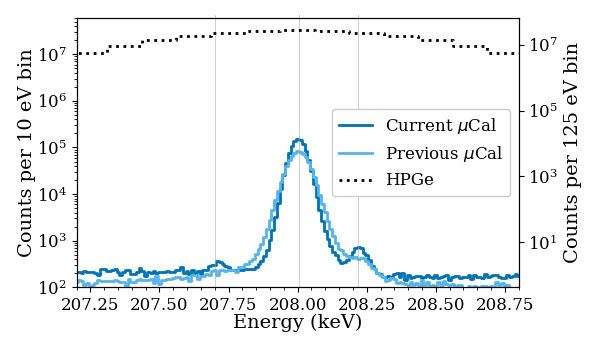}
\caption{In the microcalorimeter spectra (\textit{light blue} and \textit{blue}) of Pu standard PIDIE6-6 the ${}^{241}$Am 208.0 keV line has two satellite peaks, which are not visible in the HPGe spectrum (\textit{black dotted}). The 208.0~keV peak and the two satellite peaks are indicated with vertical grey lines. Recent advances \cite{Becker2019,Yoho2020_umatch} in microcalorimeter instrumentation and data analysis algorithms make the distinctions even sharper between previous (\textit{light blue}) and current (\textit{blue}) microcalorimeter data.}
\label{fig:spectrum}
\end{figure}

A 208~keV peak visible in Pu gamma-ray spectra is populated from the decay of both ${}^{241}$Am and ${}^{241}$Pu. ${}^{241}$Am alpha decays (100\%) to ${}^{237}$Np and 
the 208.005~keV transition ($J^{\pi} = {3/2}^- \to {5/2}^-$) is an M1 and E2 transition with a branching ratio of $7.91 \times 10^{-4} \pm 1.9 \times 10^{-5}$ photons per 100 decays \cite{Basunia2006_237, ENSDF}. ${}^{241}$Pu alpha decays to ${}^{237}$U with a decay mode of 2.47\%. ${}^{237}$U then beta decays (100\%) to ${}^{237}$Np with $21.2 \pm 0.3$ 208.005~keV photons per 100 decays~\cite{Basunia2006_237, ENSDF}.

As shown in Figure~\ref{fig:spectrum}, we observe new spectral lines from ${}^{241}$Am that are approximately 25 eV above and below the tabulated 208~keV peak. These new spectral lines were not previously detectable due to the absence of measurements in the 208~keV region from diffraction-based instruments that might have adequate resolution and the coarse relatively poor 500~eV full-width-at-half-maximum (FWHM) energy resolution of widely used High Purity Germanium (HPGe) detectors.  Figure~\ref{fig:spectrum} also shows an HPGe measurement of the 208~keV spectral region and the new peaks are entirely obscured. 

\begin{figure}
\centering
\includegraphics[width=\columnwidth]{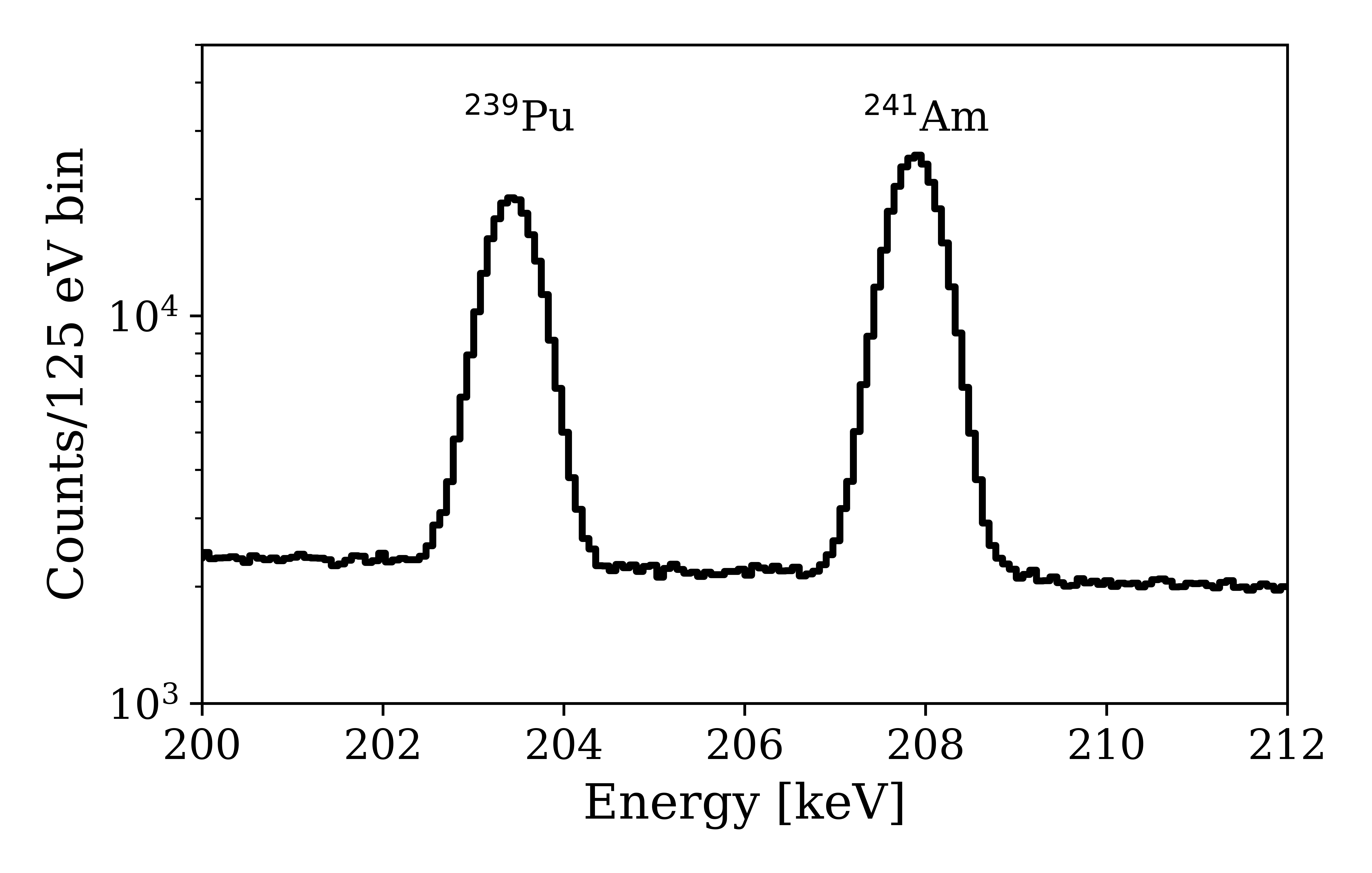}
\includegraphics[width=\columnwidth]{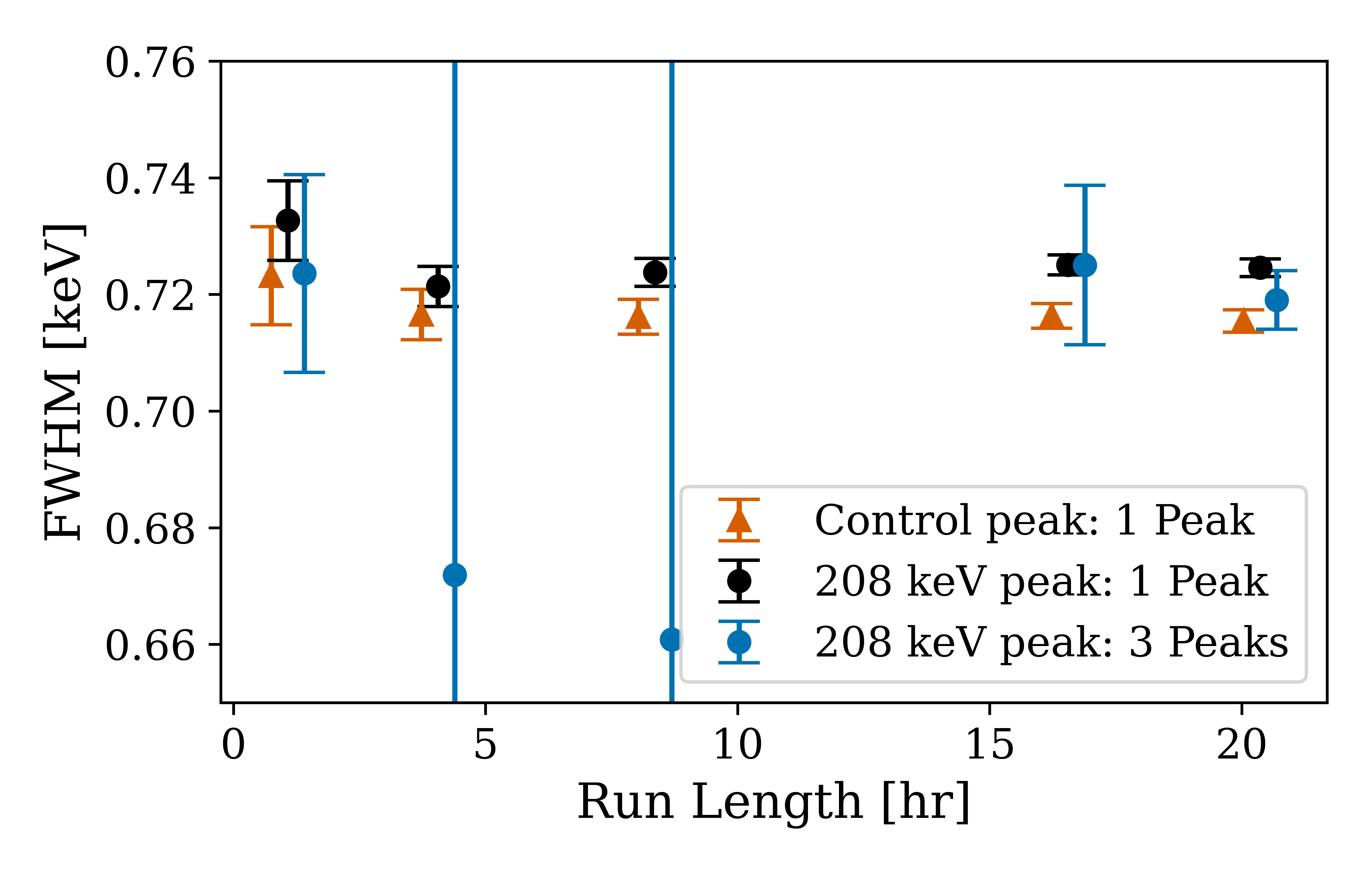}
\caption{(\textit{Top}) An HPGe measurement of Pu standard CRM138 shows a peak from ${}^{239}$Pu less than 5 keV away from the 208~keV peak, which is a result of both ${}^{241}$Am and ${}^{241}$Pu. (\textit{Bottom}) A comparison of the fit full-width-at-half-maximum (FWHM) for the control peak from ${}^{239}$Pu (\textit{orange}) to the single peak fit of the 208~keV peak (\textit{black}) shows a consistently wider FWHM on the 208~keV peak. Error bars represent 1-sigma uncertainty on the FWHM from the fitting routine. Fitting the 208-keV peak with three components (\textit{blue}) results in high uncertainty in the fit FWHM due to degeneracy of the solution.}
\label{fig:HPGe}
\end{figure}

While HPGe detectors cannot resolve the new peaks, the presence of additional peaks could perhaps have been inferred from careful study of HPGe data. In Figure~\ref{fig:HPGe}, a planar HPGe measurement of a Pu reference material (CRM138) is shown for the region near the 208~keV peak. A nearby peak from ${}^{239}$Pu at 203.54~keV allows for a peak shape comparison. As a function of measurement time, the FWHM of the ${}^{239}$Pu peak becomes increasingly differentiated from the FWHM required to fit the 208~keV peak complex. The difference in fit FWHM for the 203.54~keV peak (\textit{orange}) and the 208~keV peak complex (\textit{black}) in the last measurement of 20 hours and 22 minutes is 3.7 sigma. However, it is not possible to extract detailed information about the 208~keV peak complex from the HPGe data. For example, fitting the complex with three Gaussians whose initial centroid guesses are taken from the microcalorimeter measurements (\textit{blue}) fails to produce convergence on the known shape of the line complex. This failure shows the necessity of the high resolution gamma microcalorimeter system for understanding the details of the 208~keV peak complex.

With a recently commissioned spectrometer, SOFIA (Spectrometer Optimized for Facility Integrated Applications), measurements of Pu and Am samples showed two previously unknown gamma peaks at 207.72 and 208.21~keV. SOFIA is a 256-pixel superconducting transition-edge sensor microcalorimeter array combined with high-bandwidth microwave frequency-division multiplexed readout, providing high efficiency and high count rate capability \cite{Croce2019_INMM}. These capabilities combined with improved analysis algorithms for co-adding pixel data allowed these peaks to be clearly resolved.

In this work, SOFIA used the SLEDGEHAMMER (Spectrometer to Leverage Extensive Development of Gamma-ray TESs for Huge Arrays using Microwave Multiplexed Enabled Readout) detector array, the first large microwave-multiplexed gamma microcalorimeter array~\cite{Croce2019_INMM}. SLEDGEHAMMER was operated in multiple cryostats before it was installed in SOFIA and is well characterized. Typical spectral resolution with SOFIA is 73~eV FWHM at 208.0~keV. In a microcalorimeter, deposited energy is transduced first to a temperature change and then to an electrical signal.  The energy resolution of individual microcalorimeters is set by power fluctuations between the pixel and the heat bath, a phenomenon that has more favorable fundamental limits than counting charge in a semiconductor such as HPGe \cite{Ullom2015}. 

We acquired microcalorimeter spectra from five materials that contain various Pu isotopes and ${}^{241}$Am as well as from a sixth sample of nearly pure ${}^{241}$Am (details in Table~\ref{tab:pustandards}).  In all spectra, 1.5~mm of Cd was placed between the source and microcalorimeter array in order to attenuate the high intensity 59.54~keV ${}^{241}$Am spectral feature. Spectral features show up at 207.7~keV and 208.2~keV in all plutonium spectra, and become more prominent as the ratio of ${}^{241}$Am/${}^{241}$Pu increases. The features can be clearly seen in PIDIE-6 (reactor grade plutonium with a large percentage of ${}^{241}$Am) shown in Figure~\ref{fig:spectrum}. 

Although SOFIA has 256 pixels, only half (128) were run due to readout limitations at the time of acquisition. In order to get the best quality data, additional pixels were removed from analysis for having too low a resolution, for not enough pulses surviving cuts \cite{Becker2019}, and for being poorly co-added in the final spectrum (i.e. not all peaks were aligned between a single pixel and the final co-added spectrum). The spectrum from each pixel is co-added with a master pixel (chosen primarily for its linearity) using a cubic spline transformation to match all peaks in the spectrum above a given threshold~\cite{Yoho2020_umatch}. This differs from previous approaches in pixel co-adding which relied on co-adding energy-calibrated spectra, where energy calibration only used $9--13$ calibration points. Peak shapes from this new method of co-adding have been shown to be more uniform and Gaussian (Figure \ref{fig:spectrum}, \textit{blue}) than peak shapes from the prior co-adding method (Figure \ref{fig:spectrum}, \textit{light blue}). In the energy-calibration-first method of co-adding, peak centroids may shift by up to 10~eV from known values accompanied by slight peak shape degradation~\cite{Hoover2013, Winkler2015}. Even a small peak shape degradation is enough to make these small ${}^{241}$Am peaks disappear into the background, which is why this research team has not observed these peaks until now.

\section{Sn X-ray Escape Hypothesis}
\begin{figure}
  \centering
  \begin{minipage}[b]{\columnwidth}
  \includegraphics[width=\columnwidth]{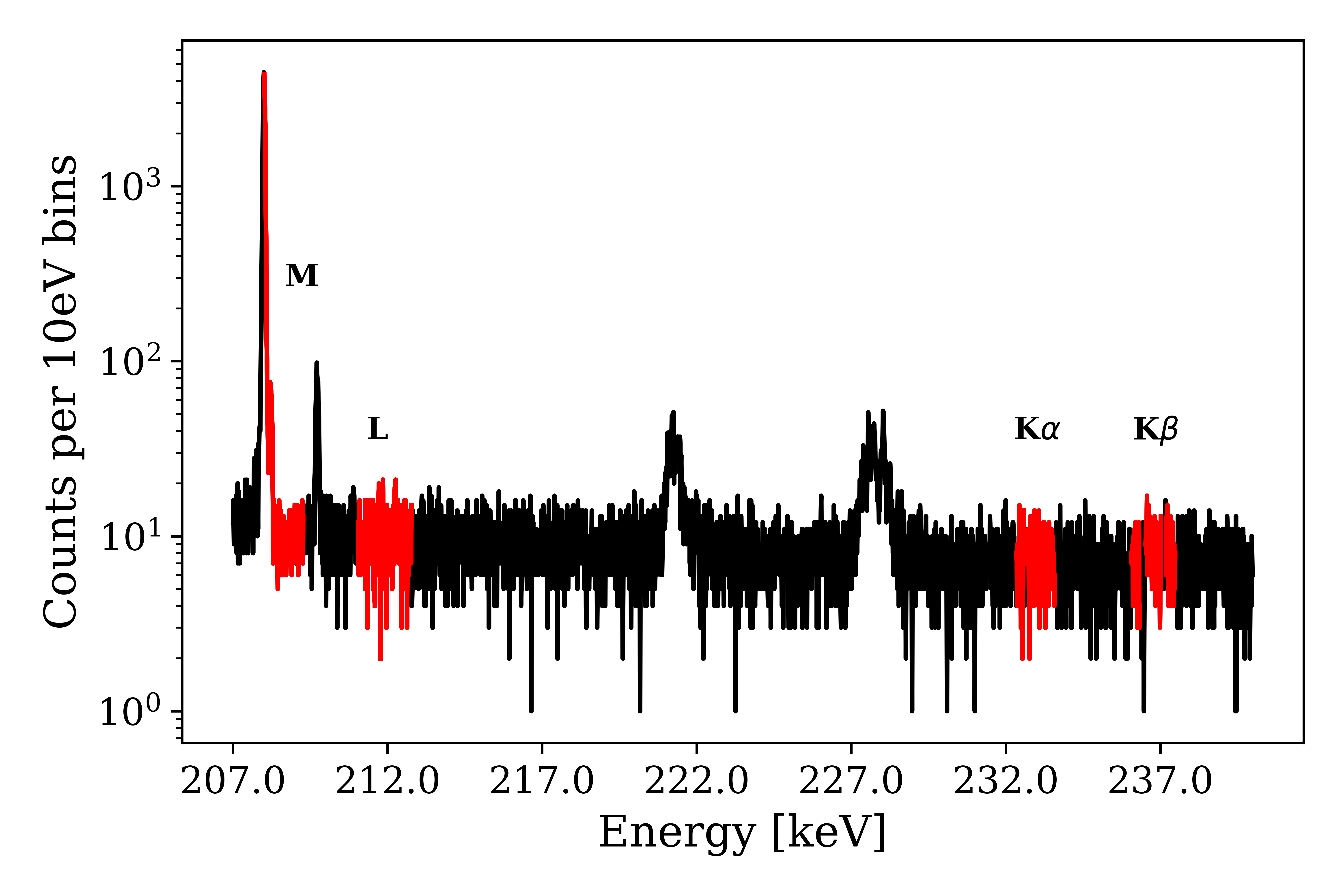}
  \end{minipage}
  \begin{minipage}[b]{\columnwidth}
  \includegraphics[width=\columnwidth]{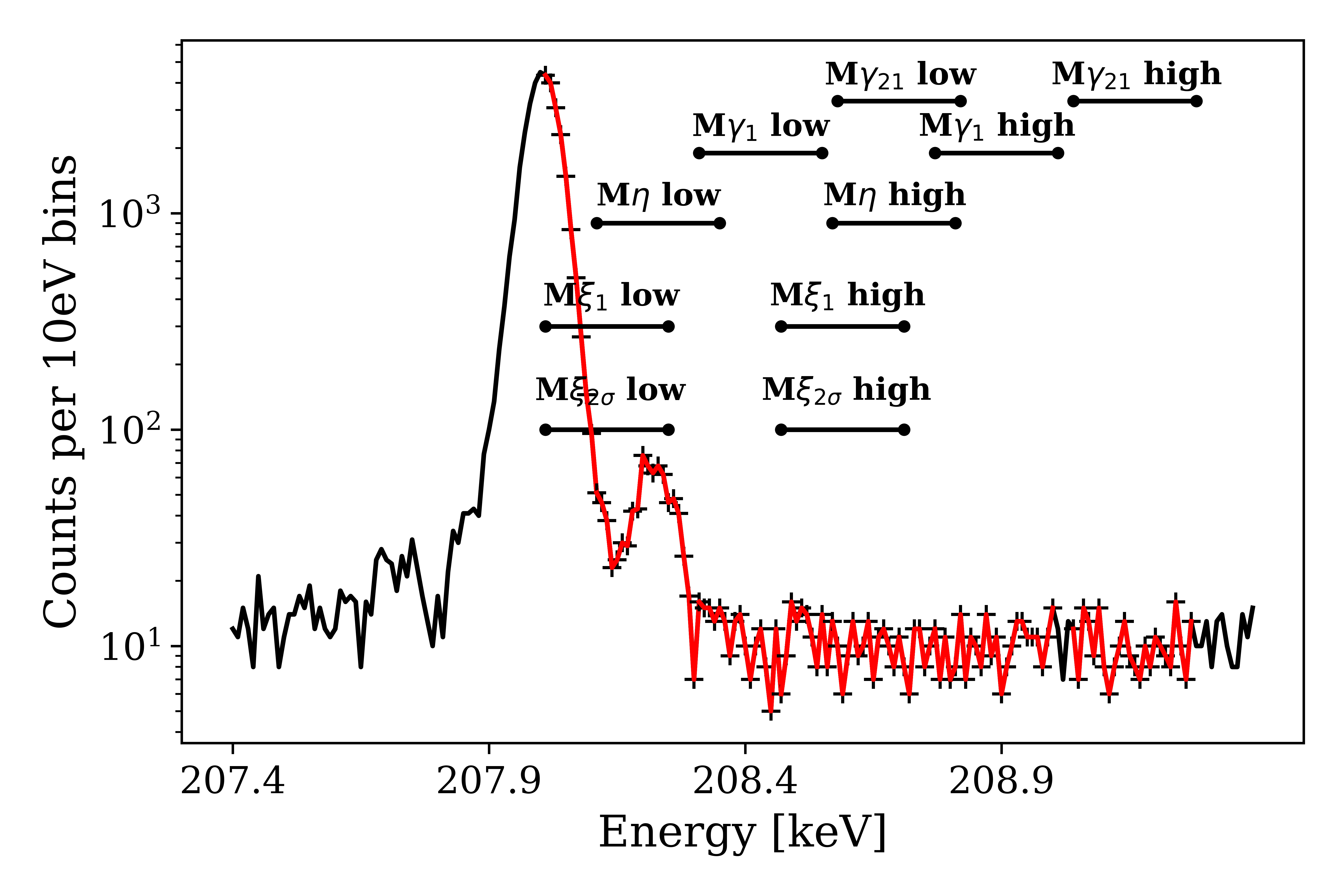}
  \end{minipage}
\caption{(\textit{Top}) Pure ${}^{241}$Am spectrum in the 208--240~keV range with regions from which a Sn X-ray could escape labeled in red. For example, the L region defines the region where a peak would have to exist, for an L X-ray to be responsible for the counts in the satellite peaks around the 208.0~keV peak. (\textit{Bottom}) Pure ${}^{241}$Am spectrum with regions from which a Sn M X-ray could escape labeled in red.}
\label{fig:spectrumwescapes}
\end{figure}

One competing explanation for the 207.7~keV and 208.2~keV peaks is that they are Sn escape peaks, a detector artifact. Here we carefully refute that hypothesis. A gamma ray (with energy $E_{\gamma}$) interacting within the microcalorimeter Sn absorber has a probability of creating a Sn X-ray (with energy $E_x$), which may escape the Sn absorber resulting in a thermalized energy of $E_{\gamma}-E_x$. The probability of this happening is referred to as the X-ray escape fraction and defined as $N_{\mathrm{escape}}/(N_{\gamma}+N_{\mathrm{escape}})$, where $N_{\mathrm{escape}}$ is the number of counts in the escape X-ray peak (at energy $E_{\gamma}-E_x$) and $N_{\gamma}$ is the number of counts in the full energy photopeak (at energy $E_{\gamma}$). For the 59.54~keV ${}^{241}$Am peak, the total escape fraction from all K Sn X-rays was previously measured to be $(25.9\pm 0.5)\%$~\cite{Hoover2009}, and the total escape fraction from all K Sn X-rays from the 208.0~keV peak was 17.5\%~\cite{Yoho2020_BR}. The most probable X-ray escape comes from the higher energy X-rays, like the K$\alpha_1$ and K$\alpha_2$ X-rays, which together account for 78--83\% of the escape probability~\cite{Hoover2009, Yoho2020_BR}. If the L X-rays escape with equal probability as the K X-rays (which they do not, since they are lower in energy), the total escape fraction cannot be more than 34\%. To address the possibility of the satellites of the 208.0~keV peak being a result of Sn escape events, all above-background counts are summed in regions corresponding to two FWHM around the sum of 207.7~keV (or 208.2~keV) and the energy of a Sn X-ray. These regions are highlighted in Figure~\ref{fig:spectrumwescapes}.

\mycomment{
\begin{table}[ht]
\caption{Counts within the satellite peaks and for the regions from which a Sn X-ray could escape to create the satellite peaks. Escape probabilities are calculated assuming the counts in the satellite peaks are counts from higher energy regions that have Sn X-rays escaping the absorber. \label{tab:escape}}
\centering
\begin{tabular}{lcc}
\toprule
	&	207.7~keV	&	208.2~keV	\\
\midrule
Counts in Peak	&	$98 \pm 15$	&	$444 \pm 24$	\\
Counts (L and K region)	&	$1 \pm 71$	&	$289 \pm 72$	\\
L+K Escape Probability	&	99\%	&	61\%	\\
Counts (M region)	&	$27075 \pm 169$	&	$-69 \pm 26$	\\
M Escape Probability	&	0.3\%	&	100\%	\\
\bottomrule
\end{tabular}
\end{table}
}

Using the pure ${}^{241}$Am spectrum, the counts in the 207.7~keV peak and the 208.2~keV peak are $98 \pm 15$	and $444 \pm 24$, respectively. The above-background area from the regions labeled L, K$\alpha$, and K$\beta$ in Figure~\ref{fig:spectrumwescapes} (\textit{Top}) are calculated. If we assume that the satellite peaks are escape peaks, the combined L and K escape probability would have to be 99\% and 61\% for the 207.7~keV peak and the 208.2~keV peak respectively. This is far higher than is reasonable given the prior measurements of escape probabilities. Let us now consider that these peaks may be a result of an M X-ray escape.

Figure~\ref{fig:spectrumwescapes} (\textit{Bottom}) shows the regions from which a Sn M X-ray would have to escape in order to create the satellite features. For example, if the high energy satellite peak were created from an escape of an M$\gamma_{21}$ X-ray, the full energy peak without the escape would appear at the region labeled ``M$\gamma_{21}$ high''. Similarly, if the low energy satellite peak were created from an escape of an M$\gamma_{21}$ X-ray, the full energy peak without the escape would appear at the region labeled ``M$\gamma_{21}$ low''. Assuming the satellites are M X-ray escapes, this would only be reasonable for the 207.7~keV peak since the required escape probability for the 208.2~keV peak to be an escape peak is too high (100\%). If all the counts in the 207.7~keV peak are M X-ray escapes, the M X-ray escape fraction is $(0.3 \pm 0.06)\%$. This value is only half the escape probability of previously measured K$\beta_2$ X-ray escapes in Sn~\cite{Hoover2009, Yoho2020_BR} and  no other high intensity peaks show this low energy satellite from an M X-ray escape peak corresponding to this escape fraction. Additionally, the 207.7 features is not the right distance from the centroid of the 208.0~keV peak. The counts in the M region are off center from the large 208.0~keV peak. Thus we conclude that neither the 207.7~keV peak nor the 208.2~keV peak are the products of Sn X-ray escapes.

\section{Nuclear Transition Possibilities}
De-excitation of the ${}^{241}$Am alpha-decay daughter product, ${}^{237}$Np, proceeds via gamma emission or conversion electrons. There are several combinations of ${}^{237}$Np levels whose energy differences agree within uncertainty with the measured centroids at 207.72(2)~keV and 208.21(1)~keV. Table~\ref{tab:evidence} shows this agreement between experimentally determined centroids and proposed photon emission energies. Propagated uncertainties in Table~\ref{tab:evidence} consist of centroid least-squares fit uncertainties and the uncertainties in the tabulated ${}^{237}$Np nuclear level energies.

\begin{table*}[ht!]
\caption{Evidence for ${}^{241}$Am assignation of the 207.7~keV and 208.2~keV peaks. The centroid is experimentally determined from the peak fits. The energy levels of the initial ($i$) and final ($f$) state come from literature and the difference is the theoretical transition energy, $E_\gamma$. The difference between the centroid and the theoretical transition energy is given in units of $\sigma$, where $\sigma$ includes the experimental uncertainty and the uncertainty of the theoretical transition added in quadrature. The measured branching ratio (BR$_m$) is per 100 disintegrations and comes from a combined analysis of pure $^{241}$Am source and Pu standards following two methods as described in the text. The probability of an ${}^{241}$Am alpha decay to populate an energy level at $E_i$ or higher is larger than the branching ratio of the transition plus the branching ratios of all the transitions from the $E_i$ state ($\Sigma I_\gamma$). Uncertainties in parentheses represent 67\% confidence intervals. \label{tab:evidence}}
\centering
\footnotesize
\begin{tabular}{rrrr}
\toprule
Centroid [keV] 	&	207.72(2) 	&	207.72(2)       	&	208.21(1)       	\\
\midrule
Transition ($J^\pi$)    	&	     \hspace{0.25in} $({7/2}^+,{9/2}^+) \to {11/2}^+$        	&	 \hspace{0.25in}${5/2}^- \to ({3/2}^-)$       	&	 \hspace{0.25in}$({11/2}^-) \to {11/2}^-$     	\\
\midrule
$E_i$ [keV]	&	805.77(12)     	&	721.96(1)      	&	434.12(5)      	\\
$E_f$ [keV]	&	597.99(9)      	&	514.19(4)      	&	225.96(2)      	\\
$E_\gamma$ [keV]        	&	207.78(15)     	&	207.77(4)      	&	208.16(5)      	\\
$\Delta E$ $(\sigma)$ 	&	-0.4	&	-1.1	&	0.9	\\
\midrule
$\text{BR}_m$ (Method 1) 	&	 $3.22(67) \times 10^{-6}$\% 	&	 $3.22(67) \times 10^{-6}$\%	&	 $1.30(18) \times 10^{-5}$\%	\\
$\text{BR}_m$ (Method 2) 	&	 $3.20(67) \times 10^{-6}$\% 	&	 $3.20(67) \times 10^{-6}$\%	&	 $1.29(19) \times 10^{-5}$\%	\\
${}^{241}$Am BR to $E_i$	&		&	0.0007	&	0.0004	\\
${}^{241}$Am BR to $E_i$ and above	&		&	$7.96 \times 10^{-4}$	&	$2.60  \times 10^{-3}$	\\
$\Sigma I_\gamma$ from $E_i$ (ENSDF)	&		&	$6.53 \times  10^{-4}$	&	$1.37  \times 10^{-5}$	\\
$\Sigma I_\gamma$ from $E_i$ (DDEP)	&		&	$7.49  \times 10^{-4}$	&	$6.43  \times 10^{-5}$	\\
\bottomrule
\end{tabular}
\end{table*}

Based on the energy levels of ${}^{237}$Np~\cite{Basunia2006_237,ENSDF}, there is only one candidate for a transition that could produce a 208.2~keV photon. The transition from 434.12~keV ($J^{\pi}=({11/2}^-)$) to 225.96~keV ($J^{\pi}={11/2}^-$) is predominantly a magnetic dipole (M1) and electric quadrupole (E2) transition. The parentheticals on the spin-parity description of the 434.12~keV state indicate a tentative spin-parity assignment.

There are three candidates for the transition producing a 207.7~keV photon. The transition 799.82~keV ($J^{\pi}={9/2}^-$) to 592.33~keV ($J^{\pi}={13/2}^+$) can be eliminated because the transition energy is not well-matched to the experimentally measured value. The transition from 721.96~keV ($J^{\pi}={5/2}^-$) to 514.19~keV ($J^{\pi}=({3/2}^-)$) would have a transition energy of 207.77~keV and is predominantly a magnetic-dipole (M1) and electric quadrupole (E2) transition, which is not unreasonable given the low intensity of this peak. The transition from 805.77~keV ($J^{\pi}=({7/2}^+,{9/2}^+)$) to 597.99~keV ($J^{\pi}={11/2}^+$) with an energy difference of 207.78~keV would predominantly be an M1 or M2 transition, depending on the actual spin and parity of the 805.77~keV energy level. A quadrupole transition is much less likely than a dipole transition, but neither are unreasonable given the low experimental branching ratio of this peak~\cite{Krane1987}. Based on this, there is one possible transition candidate for the 208.2~keV peak and two candidates for the 207.7~keV peak.

\section{Branching Ratios}
\begin{figure}[h!]
  \centering
  \begin{minipage}[b]{\columnwidth}
  \includegraphics[width=\columnwidth]{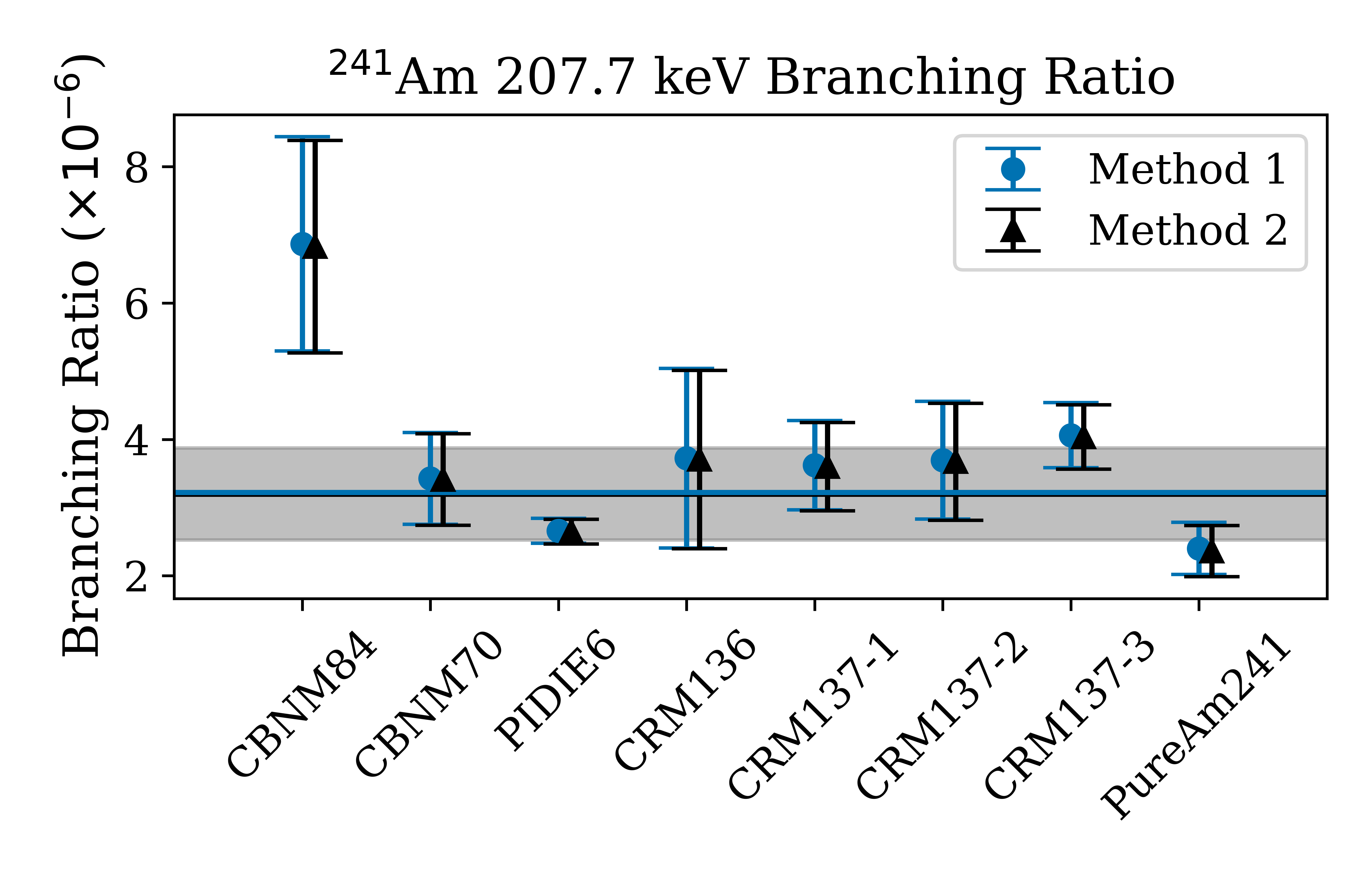}
  \end{minipage}
  \begin{minipage}[b]{\columnwidth}
  \includegraphics[width=\columnwidth]{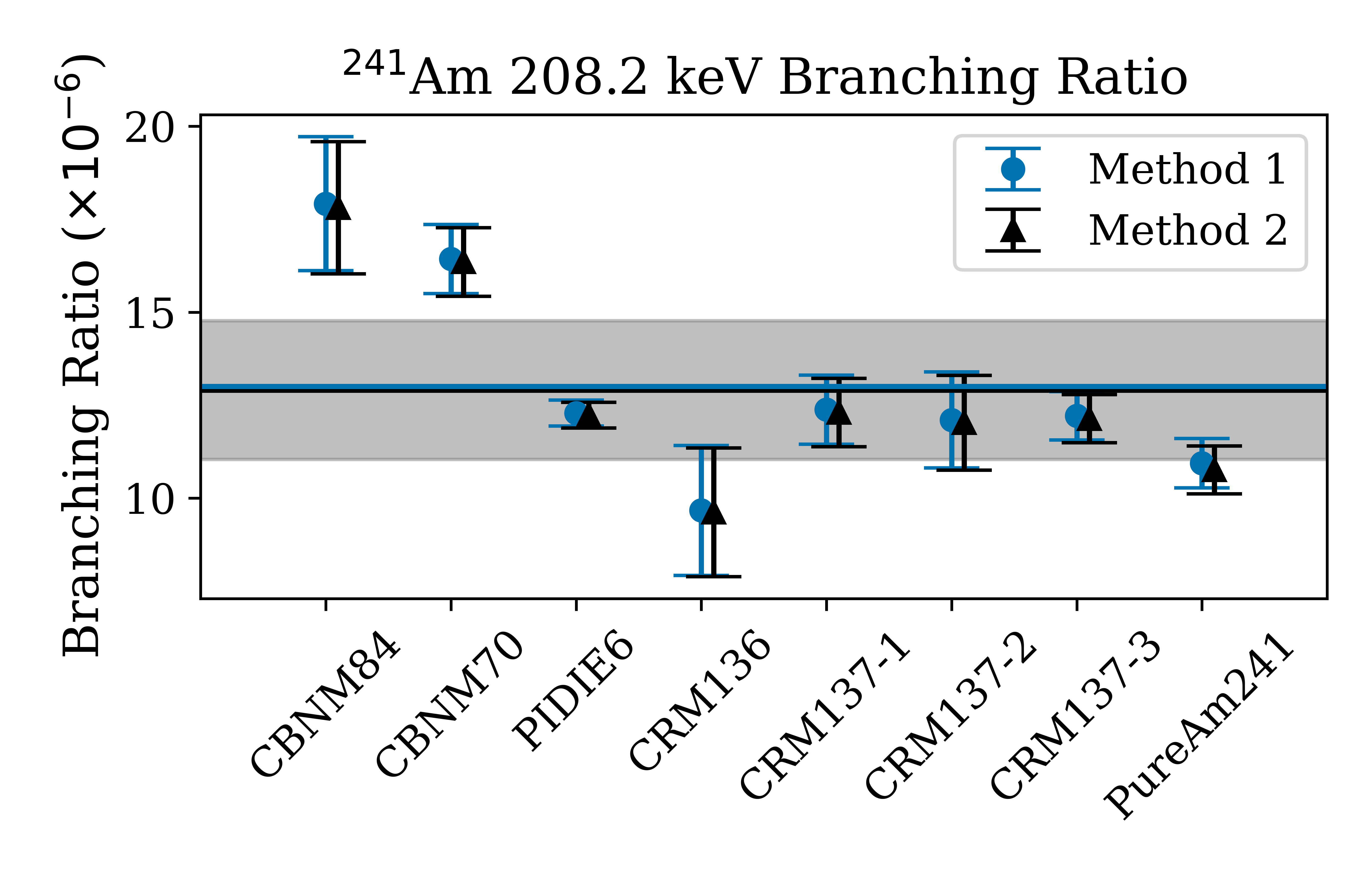}
  \end{minipage}
\caption{Branching ratios for the satellite peaks calculated from various Pu standards and a pure ${}^{241}$Am source. For both the 207.7~keV satellite peak (\textit{top}) and the 208.2~keV satellite peak (\textit{bottom}), the branching ratio was calculated in two ways: 1) assuming the ${}^{241}$Am 208.0~keV tabulated branching ratio was the actual ${}^{241}$Am 208.0~keV branching ratio (\textit{blue}) and 2) assuming the ${}^{241}$Am 208.0~keV tabulated value was the sum of the ${}^{241}$Am branching ratios for the three peaks in the 208~keV complex (\textit{black}). The three CRM137 values correspond to the three measurement dates in Table~\ref{tab:pustandards}. The best combined value from the individual spectra is given as a horizontal line. This value is calculated using the combined Type B on Bias methodology for mixed measurements. All uncertainties (including the shaded band on the best combined value) represent 67\% confidence intervals and include statistical uncertainty, reported certificate uncertainties in isotopic composition, and systematic uncertainty from half-lives and branching ratios. Fits for each of these spectra are provided in Supplmentary Material.}
\label{fig:weightedsum}
\end{figure}

Branching fractions were determined using the measured net peak areas at 207.7~keV, 208.0~keV, and 208.2~keV from several Pu standards such as those used for recent Pu and Am branching ratio measurements~\cite{Yoho2020_BR} as well as a combined sample with three ${}^{241}$Am sources, referred to as a pure ${}^{241}$Am source although it is known to contain ${}^{243}$Am. The isotopic compositions of these sources (see Table~\ref{tab:pustandards}) are used to determine the percent of the counts in the 208.0~keV peak coming from ${}^{241}$Am as opposed to ${}^{241}$Pu. Each of the three peaks is fit with two gaussians---one representing the single pixel response and one representing the effect of uncorrected drift and co-adding uncertainty---and a background step function. Similar double-Gaussian fitting functions have been successfully used in microcalorimeter spectra such as in~\cite{Hoover2013}. The branching ratio calculation is done in two ways: 1) assuming the tabulated ${}^{241}$Am branching fraction at 208.0~keV of $7.91(19) \times 10^{-4}$ photons per 100 disintegrations is the correction branching fraction for this central peak in the 208~keV complex and 2) assuming the tabulated ${}^{241}$Am branching fraction at 208.0~keV of $7.91(19) \times 10^{-4}$ photons per 100 disintegrations in the sum of the branching ratios of the three ${}^{241}$Am peaks in the 208~keV complex~\cite{ENSDF}. All calculations also require the ${}^{241}$Pu branching ratio (combined alpha-rate and ${}^{237}$U branching ratio) of $5.24(74) \times 10^{-4}$~\cite{ENSDF}. Each of these determinations and their combined value with uncertainty is given in Figure~\ref{fig:weightedsum}. The ${}^{241}$Am content for the two CRM samples is taken from a recent intercomparison exercise analysis of certified reference materials~\cite{Mathew2019}, whereas other values come from the certificate values (given in Table~\ref{tab:pustandards}). Uncertainties (67\% confidence intervals) taken into account include declared ${}^{241}$Pu and ${}^{241}$Am mass fractions for Pu items, ${}^{241}$Pu and ${}^{241}$Am branching fractions at 208.0~keV, half-lives, and net peak area least-squares fit uncertainties. A mixed method Type B on Bias methodology with a normal distribution was used to determine the best value for the branching ratio and combined uncertainty~\cite{Levenson2000,GUM2008}. In this case, each material type was treated as a separate methodology.

For method 1, the branching ratio of the 207.7~keV peak was determined to be $3.22(67) \times 10^{-6}$ photons per 100 disintegrations and the 208.2~keV peak was determined to be $1.30(18) \times 10^{-5}$ photons per 100 disintegrations. For method 2, the branching ratio of the 207.7~keV peak was determined to be $3.20(67) \times 10^{-6}$ photons per 100 disintegrations and the 208.2~keV peak was determined to be $1.29(19) \times 10^{-5}$ photons per 100 disintegrations. This second method also results in a new branching ratio for the ${}^{241}$Am 208.0~keV peak of $7.75(13) \times 10^{-4}$ photons per 100 disintegrations, compared to the original $7.91(19) \times 10^{-4}$ photons per 100 disintegrations.

Table~\ref{tab:evidence} gives the combined branching ratio from Figure~\ref{fig:weightedsum} for each peak and shows that ${}^{241}$Am decays more than often enough to elevated states of ${}^{237}$Np to account for these branching fractions. Branching ratios from both ENSDF~\cite{Basunia2006_237,ENSDF} and DDEP databases~\cite{DDEP_vol5,DDEP} are compared. The probabilities of transitioning from the $E_i$ state ($\Sigma I_\gamma$) with the additional probability from the proposed new transition ($BR_m$) is still smaller than the probability of an ${}^{241}$Am alpha decay to populate an energy level at $E_i$ or higher, so this transition is not ruled out on this basis.


\begin{table*}[!htbp]
\begin{threeparttable}
\caption{Plutonium standards and pure ${}^{241}$Am source with varying mass and composition. Mass fractions on certificate date are given in percents with respect to total plutonium. Am\% refers to the percent of counts in the 208.0~keV peak corresponding to ${}^{241}$Am on the measurement date and the total counts in the spectrum is $N$. Uncertainties in parentheses represent 67\% confidence intervals.\label{tab:pustandards}}
\footnotesize
\begin{tabular}{lllllllllll}
\toprule
Item	&	Mass [g]	&	Cert. date &	\textsuperscript{238}Pu	&	\textsuperscript{239}Pu	&	\textsuperscript{240}Pu	&	\textsuperscript{241}Pu	&	\textsuperscript{241}Am  & Meas. date & $N$ ($\times 10^6$) & Am\%\\
\midrule
CBNM84	&	6.6 oxide	&	20-06-1986 &	0.0703(3)	&	84.338(4)	&	14.207(4)	&	1.0275(9)	&	0.217(1) & 31-08-19 &8.6 & 19.5(4)	 \\
CBNM70	&	6.6 oxide	&	20-06-1986 & 0.8458(9)	&	73.319(5)	&	18.295(4)	&	5.463(2)	&	1.171(6) & 02-09-19 & 14.5  &	19.6(4) \\
PIDIE6	&	0.5 oxide	&	01-01-1988 &	0.930(6)	&	66.34(1)	&	23.89(1)	&	5.28(2)	&	3.8(2)	& 10-01-19 & 99.8 & 24.3(6)\\
CRM136	&	0.250 sulfate	&	01-10-1987 &	0.222(4)	&	84.925(8)	&	12.366(8)	&	1.902(3)	& 7.92(4)* &	17-10-18 & 14 &	31.1(6)\\
CRM137	&	0.250 sulfate	&	01-10-1987 &	0.267(3)	&	77.55(1)	&	18.79(1)	&	2.168(3)	&	7.71(6)* & 19-09-19 & 20 &	31.5(6) \\
&	&	&	&	&	&	& &	02-10-19 & 18 &	31.6(6)\\
&	&	&	&	&	&	&	& 03-10-19 & 45 &	31.6(6)\\
\midrule
Pure ${}^{241}$Am &	&	&	&	&	&	&	& 16-03-20 & 27 & 100 \\
\bottomrule
\end{tabular}
    \begin{tablenotes}
        \scriptsize
        \item[*] The ${}^{241}$Am/${}^{241}$Pu comes from \cite{Mathew2019} and has separation date of 28-4-2016, different from the certificate value for the Pu isotopics.
    \end{tablenotes}
\end{threeparttable}
\end{table*}


\section{Conclusion}
Photon signatures have been measured in various plutonium materials. The agreement displayed in Figure~\ref{fig:weightedsum} between differing materials with significantly different ${}^{241}$Am/${}^{241}$Pu ratios and different sizes strongly implies that the newly observed spectral lines at 207.7 and 208.2~keV are not artifacts from signal processing or co-adding. Evidence has been presented that these signatures originate from ${}^{241}$Am decay. Evidence includes (1) measured branching ratios from a variety of Pu standards with varying ${}^{241}$Am/${}^{241}$Pu mass fraction that agree with the determined branching ratio from the pure Am source, (2) measured peak centroids agree within uncertainty with tabulated differences in ${}^{237}$Np nuclear energies, and (3) measured branching fractions are much lower than the total net fraction of ${}^{241}$Am decays resulting in ${}^{237}$Np excited states above the probable excited state from which the transition originated. This improved picture of the number of gamma-ray spectral features in actinide spectra near 208~keV, along with their absolute energies and branching ratios, is anticipated to improve the accuracy with which the composition of nuclear materials can be determined by gamma-ray spectroscopy.

\begin{acknowledgments}
This work was supported by the G. T. Seaborg Institute, the US Department of Energy (DOE) Nuclear Energy's Fuel Cycle Research and Development (FCR\&D), Materials Protection, Accounting and Control Technologies (MPACT) Campaign and Nuclear Energy University Program (NEUP), and the NIST Innovations in Measurement Science program and is published as LA-UR-24-28099.
\end{acknowledgments}

\section*{{CRediT} authorship contribution statement}
\textbf{Katrina Koehler}: Methodology, Software, Validation, Formal Analysis, Investigation, Resources, Data Curation, Writing - Original Draft, Writing - Review \& Editing, Visualization, 
\textbf{Michael Yoho}: Conceptualization, Methodology, Software, Formal Analysis, Investigation, Data Curation, Writing - Original Draft, 
\textbf{Matthew Carpenter}: Resources, 
\textbf{Mark Croce}: Investigation, Resources, Supervision, Project Administration, Funding Acquisition, 
\textbf{David Mercer}: Supervision, 
\textbf{Chandler Smith}: Investigation, 
\textbf{Aidan Tollefson}: Investigation, Visualization, 
\textbf{Duc Vo}: Supervision, Funding Acquisition, 
\textbf{Michael Famiano}: Supervision, 
\textbf{Caroline Nesaraja}: Writing - Review \& Editing, Supervision, 
\textbf{Daniel Becker}: Resources, Writing - Review \& Editing, Supervision, 
\textbf{Johnathon Gard}: Resources, 
\textbf{Abigail Wessels}: Resources, 
\textbf{Joel Ullom}: Resources, Writing - Review \& Editing, Supervision, Funding Acquisition, 
\textbf{Douglas Bennett}: Resources, 
\textbf{John Mates}: Resources, 
\textbf{Nathan Ortiz}: Resources, 
\textbf{Daniel Schmidt}: Resources, 
\textbf{Daniel Swetz}: Resources, 
\textbf{Leila Vale}: Resources


%

\end{document}